\documentclass{ifacconf}
\usepackage{algorithm}
\usepackage{algorithmic}
\usepackage{amsmath}
\usepackage{amssymb}
\usepackage{graphicx}
\usepackage[subfigure]{graphfig}
\usepackage{subfigure}
\usepackage{multirow}
\usepackage{float}
\usepackage{booktabs}
\usepackage{mathrsfs}
\usepackage[numbers,sort&compress]{natbib}
\makeatletter

\newcommand{\Rmnum}[1]{\expandafter\@slowromancap\romannumeral #1@}
\makeatother
\makeatletter
\def\hlinew#1{%
	\noalign{\ifnum0=`}\fi\hrule \@height #1 \futurelet
	\reserved@a\@xhline}
\makeatother          
\begin{document}
	\begin{frontmatter}
		
		\title{A New Fuzzy $H_{\infty}$ Filter Design for Nonlinear Time-Delay Systems with Mismatched Premise Membership Functions\thanksref{footnoteinfo}} 
		
		\thanks[footnoteinfo]{This study was partly supported by the National Natural Science Foundation of China (61304108,51307035)}
		
		\author[First]{Qianqian Ma} 
		\author[First]{Hongwei Xia} 
		\author[Second]{Lili}
		\author[First]{Guangcheng Ma}
		
		\address[First]{Control Science and Engineering Department,
			Harbin Institute of Technology, Harbin 150001, P. R. China, (email: maqq222008@hit.edu.cn)}
		
		\address[Second]{School of Information Science and Engineering,
			Harbin Institute of Technology at Weihai, Weihai 264200, P. R. China, (email: lili406@hitwh.edu.cn)}

		\begin{abstract}                
			This paper is concerned with the fuzzy $H_{\infty}$ filter design issue for nonlinear systems with time-varying delay. To overcome the shortcomings of the conventional methods with matched preconditions, the fuzzy  $H_{\infty}$ filter to be designed and the T-S fuzzy model are assumed to have different premise membership functions and number of rules, thus, greater design flexibility and robustness to uncertainty can be achieved. However, such design will also make the derived results conservative, to relax the result, a novel integral inequality which is tighter than the traditional inequalities derived from the Leibniz-Newton formula is applied, besides, a fuzzy Lypunov function and the information of the membership functions are also introduced. All the design methods are presented in LMI-based conditions. Finally, two numerical examples are given to prove the effectiveness and superiority of the proposed approach.
		\end{abstract}
		
		
	\end{frontmatter}
	
	\section{Introduction}
	It is well known that time-delay can cause instability and deteriorate the performance of the systems, also it is inevitable in practical control systems. So the study of control systems with time-delay is critical, and researchers have devoted great efforts to investigate them (\cite{analysis2000,H1998,delayt2004,Wang2020}). And in this paper, we will mainly investigate nonlinear systems with time-varying delay.
	
	Moreover, filtering technology is playing a crucial role in signal processing, and over the development of several decades, fruitful research results have been obtained (\cite{steady1989,Delay2003,A2001,kong2016temporal,kong2016new, Ma2017}), among them, $H_{\infty}$ filtering (\cite{A2001,star,Qin2021,ma2022}) has attracted widespread attention, as $H_{\infty}$ filter has no particular requirement for external noise signal and it is not sensitive to uncertainty. In the past few years, researchers has proposed various techniques to improve the performance of the $H_{\infty}$ filter, to mention a few, in (\cite{star}), a delay-dependent fuzzy $H_{\infty}$ filter design method was proposed for T-S fuzzy-model-based system with time-varying delay. However, in this paper, the Lyapunov-Krasovskii function candidate was chosen as a single Lyapunov function, to obtain more relaxed results, the literature  (\cite{JH}) adopted a fuzzy Lypunov function to analyze the stability condition. In (\cite{Y}), the fuzzy $H_{\infty}$ filter design approach was improved through estimating the upper bound of the derivative of Lyapunov function without ignoring any useful terms. On the basis of (\cite{Y}), literature (\cite{short}) proposed a technique to obtain more accurate upper bound of the derivative of Laypunov function.
	
	 However, when designing $H_{\infty}$ filter for nonlinear systems with time-delay, all the proposed methods require matching conditions, i.e, the fuzzy filter and the fuzzy model are assumed to have same membership functions, such assumption can facilitate the filter design, but on the other hand, it will also limit the design flexibility and make the designed filter lose robustness to the system with uncertain membership functions. To address this problem, we will apply the imperfect premise matching method (\cite{imperfect2009},\cite{kong2018efficient}) to design fuzzy $H_{\infty}$ filter, which means the fuzzy filter to be designed and the fuzzy model will be allowed to have different premise membership functions and number of fuzzy rules, thus successfully address the issue of traditional $H_{\infty}$ filter design methods. Nevertheless, such design will also make the designed filter tend to be more conservative. So in this paper, we will adopt a fuzzy Lyapunov function to derive the fuzzy $H_{\infty}$ filter, which can increase the number of free-weighting matrices, and relax the conservatism of the result.
	
     Besides, as $H_{\infty}$ filter has to ensure the filtering error system is asymptotically stable, the stability analysis is necessary, and in this process, the inequalities derived from the Leibniz-Newton formula is often adopted, like in literature (\cite{short,star,further2007,new2009,Fuzzy2007,Ma2017b}). Though the $H_{\infty}$ filter design issue can be solved, the derived results are conservative, and there is little room left to improve. So in this paper, we aim to use a new integral inequality to substitute the conventional inequalities derived from the Leibniz-Newton formula, and to further relax the designed fuzzy $H_{\infty}$  filter.
	
	What's more, usually the existing fuzzy $H_{\infty}$ filter design methods are membership functions independent (\cite{short,star,further2007,kong2017impact,kong2017object}), to the best of our knowledge, the membership functions dependent $H_{\infty}$ filter design approach is yet to be thoroughly investigated, so in order to further reduce the conservatism, we will consider the information of the membership functions in the criterion. 
	
	For the sake of obtaining satisfying fuzzy $H_{\infty}$  filter, we are going to employ the mismatching method to design fuzzy filter, and to relax the results, a novel integral inequality which is tighter than conventional inequalities will be applied, besides the information of the membership functions and a fuzzy Lyapunov function technique will be introduced into the criteria as well. And the rest of this paper will be organized as follows: Section 2 systematically describes the problem we are going to solve and gives a relevant lemma about the new integral inequality. In Section 3, the detailed $H_{\infty}$ design methods are presented. And Section 4 uses three simulation examples to illustrate the effectiveness of the designed methods. Finally, some conclusions are given in Section 5.
	
	\section{Preliminaries}

	Consider a nonlinear system involving time-varying delay, which is described by the following $p$-rule T-S fuzzy model:
	\subsubsection*{Plant Rule $i$}
	IF $\psi_{1}(t) $ is $\mathcal{M}_{1}^i$ and
	$\psi_{2}(t)$ is $\mathcal{M}_{2}^i$ and $\ldots$ and $\psi_{m}(t)$ is $\mathcal{M}_{m}^i$, THEN
	\begin{eqnarray}
	\dot{x}(t)&=&A_{i}x(t)+A_{\tau{i}}x(t-\tau(t))+B_{i}w(t),\notag\\
	y(t)&=&C_ix(t) + C_{\tau{i}}x(t-\tau(t) + D_iw(t),\notag\\
	z(t)&=&E_ix(t)+E_{\tau{i}}x(t-\tau(t)), \notag\\
	x(t)&=&\chi(t), \quad \forall  t\in{[-\tau_0,0]}, 
	\end{eqnarray}
	where  $i=1,2,\ldots,p$. $\psi_{\alpha}(t)(\alpha=1,2,\ldots,m)$ is the premise variable. $\mathcal{M}_{\alpha}^i$ is the fuzzy term of rule $i$ which corresponds to the function $\psi_{\alpha}$. $m$ is a positive integer. And $ x(t)\in\mathbb{R}^{n} $ is the system state, $z(t)\in \mathbb{R}^{q}$ is the unknown signal to be estimated, $y(t)\in \mathbb{R}^{m}$ is the system output, $w(t)\in\mathbb{R}^{p}$ is the noise signal which is assumed to be arbitrary and satisfy $w(t)\in L_2\in[0,\infty)$. $A_i, A_{\tau{i}}, B_i,C_i, C_{\tau{i}}, D_i, E_i,E_{\tau{i}}$ are given system matrices. Time delay $ \tau(t) $ is a continuously differentiable function, satisfying the conditions followed:
	\begin{equation}\label{tau}
	0\leq \tau(t)<h, \qquad \dot{\tau}(t)\leq \rho.
	\end{equation}
	By fuzzy blending, the system dynamics can be presented as

	\begin{equation}\label{fuzzy model}
	\begin{aligned}
	\dot{x}(t)=&\sum_{i=1}^{p}\phi_{i}(\psi(t))[A_{i}x(t)+A_{\tau{i}}x(t-\tau(t))+B_{i}w(t)],\notag\\
	y(t)=&\sum_{i=1}^{p}\phi_{i}(\psi(t))(C_ix(t) + C_{\tau{i}}x(t-\tau(t)+ D_iw(t)),\notag
	\end{aligned}
	\end{equation}
	\begin{equation}
	z(t)=\sum_{i=1}^{p}\phi_{i}(\psi(t))(E_ix(t)+E_{\tau{i}}x(t-\tau(t))).
	\end{equation}
	where
	\begin{equation}\label{w(x(t))}
	\begin{aligned}
	&\psi(t)=(\psi_1(t),\psi_2(t),\ldots,\psi_m(t)),\\
	&\sum_{i=1}^{p}\phi_{i}(\psi(t))=1, \phi_{i}(\psi(t))\geqslant0,\\
	&\phi_{i}(\psi(t))=\frac{\Pi_{\alpha=1}^{m}\mu_{\mathcal{M}_{\alpha}^i}(\psi_{\alpha}(t))}{\sum_{k=1}^{p}\Pi_{\alpha=1}^{m}\mu_{\mathcal{M}_{\alpha}^k}(\psi_{\alpha}(t))}, 	
	\end{aligned}
	\end{equation}
	and $\phi_{i}(\psi(t))$ is the normalized membership function; $\mu_{\mathcal{M}_{\alpha}^i}(\psi_{\alpha}(x(t)))
	$ is the grade of membership function which corresponds to the fuzzy term $\mathcal{M}_{\alpha}^i$.

	In order to eliminate the drawbacks of the conventional PDC methodlogy, in this paper, the fuzzy $H_{\infty}$ filter to be designed is allowed to have different premise membership functions and number of rules from the fuzzy model. Hence, we assume that the number of the fuzzy filter to be designed is $c$, and it can be described as:
	\subsubsection*{Filter Rule $j$} 
	IF $g_{1}(t) $ is $\mathcal{N}_{1}^{j}$ and
	$g_{2}(t)$ is $\mathcal{N}_{2}^{j}$ and $\ldots$ and $g_{\Theta}(t))$ is $\mathcal{N}_{\Theta}^{j}$, THEN	
	\begin{equation}
	\begin{aligned}
	\dot{x_f}(t)&=A_{fj}x_f(t)+B_{fj}y(t),\\
	z_f(t)&=C_{fj}x_f(t),
	\end{aligned}
	\end{equation}	
	where $x_f(t)\in \mathbb{R}^{n}$ and $z_f(t)\in \mathbb{R}^{q}$ are the state and output of the fuzzy $H_{\infty}$ filter respectively.  And $A_{fj},B_{fj},C_{fj}$ are the filter matrices of that will be designed. 
	
	Similarly, through fuzzy blending, the fuzzy $H_{\infty}$ filter to be designed can be presented as	
	\begin{equation}\label{fuzzyfilter}
	\begin{aligned}
	\dot{x_f}(t)&=\sum_{j=1}^{c}n_{j}(g(t))(A_{fj}x_f(t)+B_{fj}y(t)),\\
	z_f(t)&=\sum_{j=1}^{c}n_{j}(g(t))C_{fj}x_f(t),
	\end{aligned}
	\end{equation}
	where 
	\begin{equation}
	\begin{aligned}
& g(t)=(g_1(t),g_2(t),\ldots,g_{\Theta}(t)),\\
&\sum_{j=1}^{c}n_{j}(g(t))=1, n_{j}(g(t))\geqslant 0, \\
& n_j(g(t))=\frac{\Pi_{\beta=1}^{\Theta}\mu_{\mathcal{N}_{\beta}^j}(g_{\beta}(x_f(t)))}{\sum_{k=1}^{c}\Pi_{\beta=1}^{\Theta}\mu_{\mathcal{N}_{\beta}^k}(g_{\beta}(x_f(t)))},
	\end{aligned}
	\end{equation}  	
	for all $j$, $n_j(g(t))$ is the normalized membership function. $\mu_{\mathcal{N}_{\beta}^j}(g_{\beta}(x_f(t)))$( $\beta=1,2,\ldots,\Theta$) is the grade of membership functions which corresponds to the fuzzy term $\mathcal{N}_{\beta}^j$.
	
	
	
	According to (\ref{fuzzy model}) and (\ref{fuzzyfilter}), and define the augmented state vector as $\zeta(t)=[x^T(t),x_f(t)]^T$ and $e(t)=z(t)-z_f(t)$, we can obtain the $H_{\infty}$ filtering system as follows:
	\begin{equation}\label{Closedfiltersystem}
	\begin{aligned}
	& \dot{\zeta}(t)=\bar{A}(t)\zeta(t)+\bar{A}_{\tau}(t)\zeta(t-\tau(t))+\bar{B}(t)w(t)), \\
	&   	e(t)=\bar{E}(t)\zeta(t) + \bar{E}_{\tau}(t)\zeta(t-\tau(t))),	
	\end{aligned}
	\end{equation}
	where $\zeta(0)=[\chi(t), x_{f0}]$ for $\forall t \in[-\tau_0,0]$, and
	\begin{equation}
	\begin{aligned}
	&\bar{A}(t)=\sum_{i=1}^{p}\sum_{j=1}^{c}\phi_{i}(\psi(t))n_{j}(g(t))\left[
	\begin{array}{cc}
	A_i & 0\\
	B_{fj}C_i & A_{fj}
	\end{array}
	\right]\notag&
	\end{aligned}
	\end{equation}
	\begin{equation}\label{simplication}
	\begin{aligned}
	&\hspace{0.82cm}=\left[
	\begin{array}{cc}
	A_{\tau}(t) & 0\\
	B_f(t)C_{\tau}(t) & 0
	\end{array}
	\right],&\\
	&\bar{B}(t)=\sum_{i=1}^{p}\sum_{j=1}^{c}\phi_{i}(\psi(t))n_{j}(g(t))\left[
	\begin{array}{c}
	B_i\\   
	B_{fj}D_i
	\end{array}
	\right]\\
	&\hspace{0.68cm}=\left[
	\begin{array}{c}
	B(t)\\   
	B_f(t)D(t)
	\end{array}
	\right], \\	
	&\bar{E}(t)=\sum_{i=1}^{p}\sum_{j=1}^{c}\phi_{i}(\psi(t))n_{j}(g(t))\begin{matrix}
	[E_i & -C_{fj}
	\end{matrix}]&\\
	&\hspace{0.68cm}=[
	\begin{matrix}
	E(t) & -C_f(t)
	\end{matrix}], \\
	&\bar{E}_{\tau}(t)=\sum_{i=1}^{p}\sum_{j=1}^{c}\phi_{i}(\psi(t))n_{j}(g(t))[\begin{matrix}
	E_{\tau i} & 0
	\end{matrix}]=[\begin{matrix}
	E_{\tau}(t) & 0
	\end{matrix}].
	\end{aligned}
	\end{equation}
	
	Ergo, the fuzzy $H_{\infty}$ filter design issue that will be investigated in this paper can be presented as follows:
	\subsubsection*{Fuzzy $H_{\infty}$  filter issue}   Design a fuzzy filter in the form of $(\ref{fuzzyfilter})$ satisfying the following two conditions: 
	
	(1) If $w(t)=0$, the filtering system  (\ref{Closedfiltersystem}) is asymptotically stable;
	
	(2) For a given scalar $\gamma >0$, if $\zeta(t)\equiv 0$ for $t\in[-\tau_0,0]$, the following $H_\infty$ performance can be  satisfied for all the $T>0$ and $w(t) \in L_2[0,\infty)$.
	\begin{eqnarray}\label{Hinf definition}
	\int_0^T\|e(t)\|^2dt\leq \gamma^2\int_0^T\|w(t)\|^2dt.
	\end{eqnarray}

	What's more, the following lemma is useful for the later deduction of the main results.
	\subsubsection*{Lemma 1 (\cite{New2015})} It is assumed that $x$ is a differentiable function: $[\alpha,\beta]\rightarrow \mathbb{R}^{n}$. For $N_{1}, N_{2}, N_{3} \in \mathbb{R}^{4n\times n}$, and  $R \in \mathbb{R}^{n\times n }> 0$, the following inequality holds:
	\begin{equation}\label{lemma}
	-\int_{\alpha}^{\beta}\dot{x}^{T}(s)R\dot{x}(s)ds\leq \xi^{T} \Omega \xi,
	\end{equation}
	where
	\begin{align*}
	&\Omega = \tau(N_{1}R^{-1}N_{1}^{T}+\frac{1}{3}N_{2}R^{-1}N_{2}^{T}+\frac{1}{5}N_{3}R^{-1}N_{3}^{T})\\
	&\quad\quad+Sym\{N_{1}\Delta_{1}+N_{2}\Delta_{2}+N_{3}\Delta_{3}\},\\
		& d=\beta-\alpha,\\
	& e_{i}=\left[\
	\begin{matrix}
	0_{n\times(i-1)n}\quad I_{n} & 0_{n\times(4-i)n}
	\end{matrix}\right],\quad i=1,2,3,4,\\	
	&\Delta_{1}=e_{1}-e_{2},\quad \Delta_{2}=e_{1}+e_{2}-2e_{3},\\
	&\Delta_{3}=e_{1}-e_{2}-6e_{3}+6e_{4}\\
	&\xi=[\begin{matrix}
	x^{T}(\beta) & x^{T}(\alpha) & \frac{1}{d}\int_{\alpha}^{\beta}x^{T}(s)ds & \frac{2}{d^{2}}\int_{\alpha}^{\beta}\int_{\alpha}^{s}x^{T}(u)duds 
	\end{matrix}]^{T}.\\
	\end{align*} 	
	\section{Main Results}  
	As fuzzy $H_{\infty}$ filter has to guarantee the asymptotical stability of the filtering system, we will first analyze the stability of the system (\ref{Closedfiltersystem}).
	\subsubsection*{Lemma 2}
	For system (\ref{Closedfiltersystem}), the constants $h$, $\rho$ and $\gamma>0$ are prescribed, it will be asymptotically stable with $w(t)\equiv0$, and satisfy the $H_{\infty}$ performance condition (\ref{Hinf definition}), if there exist matrices $M\in\mathbb{R}^{2n\times 2n}>0$, $N(t)\in\mathbb{R}^{2n\times 2n}>0$, $O(t)\in\mathbb{R}^{2n\times 2n}>0$, such that the the following inequality is feasible.
	\begin{equation}\label{lemma1}
	\Omega(t)=\left[\begin{matrix}
	\Theta(t) & \sqrt{h}\Gamma^{T}_{1}(t)M & \Gamma^{T}_2(t)\\
	* & -MO(t)^{-1}M & 0\\
	* & * & -1
	\end{matrix}\right]<0,
	\end{equation}
	where
	\begin{flalign*}
	& \Theta(t)=(\Xi_1(t)+\Xi_2(t)+\Xi_3(t))+\Theta_3(t),\\
	&\Xi_1(t)=\frac{1}{h}\left[
	\begin{matrix}
	-O(t) & O(t) & 0 & 0 & 0\\
	* & -O(t) & 0 & 0 & 0\\
	* & * & 0 & 0 & 0\\
	* & * & * & 0 & 0\\
	* & * & * & * & 0
	\end{matrix}\right],&\\
	&\Xi_2(t)=\frac{3}{h}\left[
	\begin{matrix}
	-O(t) & -O(t) & 2O(t) & 0 & 0\\
	* & -O(t) & 2O(t) & 0 & 0\\
	* & * & -4O(t) & 0 & 0\\
	* & * & * & 0 & 0\\
	* & * & * & * & 0
	\end{matrix}\right],&\\
	&\Xi_3(t)=\frac{5}{h}\left[
	\begin{matrix}
	-O(t) & O(t) & 6O(t) & -6O(t) & 0\\
	* & -O(t) & -6O(t) & 6O(t) & 0\\
	* & * & -36O(t) & 36O(t) & 0\\
	* & * & * & -36O(t) & 0\\
	* & * & * & * & 0
	\end{matrix}\right],&\\
	&\Theta_{3}(t)=\left[\begin{matrix}
	\theta_1 & M\bar{A}_{\tau}(t) & 0 & 0 & M\bar{B}(t)\\
	* & -N(t) & 0 & 0 & 0\\
	* & * & 0 & 0 & 0\\
	* & * & * & 0 & 0\\
	* & * & * & * & -\gamma^{2}
	\end{matrix}\right],\\
	&\theta_1=Sym\{M\bar{A}(t)\}+N(t),&\\
	&\Gamma_1(t)=\left[\begin{matrix}
	\bar{A}(t) & \bar{A}_{\tau}(t) & 0 & 0 & \bar{B}(t)
	\end{matrix}\right]&\\
	&\Gamma_2(t)=\left[\begin{matrix}
	\bar{E}(t) & \bar{E}_{\tau}(t) & 0 & 0 & 0
	\end{matrix}\right].	&
	\end{flalign*}
	\subsubsection*{proof}
	To relax the result, the Lyapunov-Krasovskii functional candidate can be defined as fuzzy Lyapunov function:
	\begin{equation}\label{LKFfuzzy}
	\begin{aligned}
	V(t)=&\varepsilon^{T}(t)M\varepsilon(t)+\int_{t-\tau(t)}^{t}\varepsilon^T(s)N(s)\varepsilon(s)ds\\
	&+\int_{-h}^{0}\int_{t+\theta}^{t}\dot{\varepsilon}(s)^T O(s) \dot{\varepsilon}(s)dsd\theta,	
	\end{aligned}
	\end{equation}   
	where
	\begin{equation}\label{fuzzy}
	M>0,\quad N(t)=\sum_{i=1}^{p}\phi(\psi(t))N_i,\quad O(t)=\sum_{i=1}^{p}\phi(\psi(t))O_i.
	\end{equation}
	
	It is necessary to mention that in the Lyapunov function, the matrix $M$ is presented as constant matrix, but $N(t)$ and $O(t)$ are expressed as matrix function, this is because if matrix $M$ is denoted as matrix function, it will be difficult to determine the upper bound of $\Vert\dot{\phi}(x(t))\Vert$.
	
	Differentiating (\ref{LKFfuzzy}) along the trajectories of system (\ref{Closedfiltersystem}) yields:
	\begin{equation}\label{LKFdot}
	\begin{aligned}
	\dot{V}(t)=& 2\varepsilon^{T}(t)\dot{\varepsilon}(t)-(1-\dot{\tau}(t))\varepsilon^{T}(t-\tau(t))N(t)\varepsilon(t-\tau(t))\\
	&+h\dot{\varepsilon}(t)^TO(t)\dot{\varepsilon}(t)-\int_{t-h}^{t}\dot{\varepsilon}(s)^TO(t)\dot{\varepsilon}(s)ds.
	\end{aligned}
	\end{equation}  
	Applying Lemma 1, we can derive
	\begin{equation}
	\begin{aligned}
	&-\int_{t-h}^{t}\dot{\varepsilon}^{T}(s)O(t)\dot{\varepsilon}(s)ds\notag&\\
	&<-\int_{t-\tau(t)}^{t}\dot{\varepsilon}^{T}(s)O(t)\dot{\varepsilon}(s)ds\\	
		&\leq\xi^{T}(t)\left[\tau(t)F_{1}O(t)^{-1}F_{1}^{T}+\frac{\tau(t)}{3}F_{2}O(t)^{-1}F_{2}^{T}\right.\\
	&+\frac{\tau(t)}{5}F_{3}O(t)^{-1}F_{3}^{T}+Sym\{F_{1}\Pi_{1}+F_{2}\Pi_{2}+F_{3}\Pi_{3}\}\bigg]\xi(t)&\\	
	&<\xi^{T}(t)\left[hF_{1}O(t)^{-1}F_{1}^{T}+\frac{h}{3}F_{2}O(t)^{-1}F_{2}^{T}+\frac{h}{5}F_{3}O(t)^{-1}F_{3}^{T}\right.&\\
	&\quad+Sym\{F_{1}\Pi_{1}+F_{2}\Pi_{2}+F_{3}\Pi_{3}\}\bigg]\xi(t)&\\
	&=\xi^{T}(t)\left(\Theta_{1}(t)+\Theta_{2}\right)\xi(t),&
	\end{aligned}
	\end{equation}
	where 
	\begin{flalign*}
	\begin{aligned}
	&\xi(t)=[\begin{matrix}
	\varepsilon^{T}(t) & \varepsilon^{T}(t-\tau(t)) & \frac{1}{\tau(t)}\int_{t-\tau(t)}^{t}\varepsilon^{T}(s)ds & \theta_{2} & w(t)
	\end{matrix}]^{T},\\
	& \theta_2=\frac{2}{\tau^{2}(t)}\int_{t-\tau(t)}^{t}\int_{t-\tau(t)}^{s}\varepsilon^{T}(u)duds, \\
	&\Theta_{1}(t)=hF_{1}O(t)^{-1}F_{1}^{T}+\frac{h}{3}F_{2}O(t)^{-1}F_{2}^{T}+\frac{h}{5}F_{3}O(t)^{-1}F_{3}^{T}&\\
	&\Theta_{2}=Sym\{F_{1}\Pi_{1}+F_{2}\Pi_{2}+F_{3}\Pi_{3}\}&\\
	& e_{i}=\left[\
	\begin{matrix}
	0_{2n\times(i-1)2n}\quad I_{2n} & 0_{2n\times(5-i)2n}
	\end{matrix}\right],\quad i=1,2,3,4,5,\\	
	&\Pi_{1}=e_{1}-e_{2},\quad \Pi_{2}=e_{1}+e_{2}-2e_{3},\\
	&\Pi_{3}=e_{1}-e_{2}-6e_{3}+6e_{4}.\\
	\end{aligned}
	\end{flalign*}
	
	Based on inequality $(\ref{tau})$ and equation (\ref{Closedfiltersystem}), we can derive
	\begin{equation}
	\begin{aligned}
	\dot{V}(t)&< 2\varepsilon^{T}(t)M[\bar{A}(t)\varepsilon(t)+\bar{A}_{\tau}(t)\varepsilon(t-\tau(t))+\bar{B}(t)w(t)]\\
	&\quad-(1-\rho)\varepsilon^{T}(t-\tau(t))N(t)\varepsilon(t-\tau(t))\\
	&\quad+h\dot{\varepsilon}(t)^TO(t)\dot{\varepsilon}(t)+\xi^{T}(t)(\Theta_{1}(t)+\Theta_{2})\xi(t),
	\end{aligned}
	\end{equation}
	and through a straightforward computation we can obtain:
	\begin{equation}\label{deduction}
	\begin{aligned}
	&\dot{V}(t)+e^{T}(t)e(t)-\gamma^{2}w^{T}(t)w(t)<\xi^{T}(t)(\Theta_1(t)+\Theta_2\\
	&+\Theta_3(t)+h\Gamma^{T}_1O(t)\Gamma_1(t)+\Gamma^{T}_2\Gamma_2(t))\xi(t),
	\end{aligned}
	\end{equation}
	where $\Theta_3(t)$, $\Gamma_1(t)$, $\Gamma_2(t)$ are defined in $(\ref{lemma1})$.
	
	From the inequality (\ref{deduction}), it can be inferred that if
	\begin{equation}\label{transfer}
	\Theta_1(t)+\Theta_2+\Theta_3(t)+h\Gamma^{T}_1O(t)\Gamma_1(t)+\Gamma^{T}_2\Gamma_2(t)<0,
	\end{equation}
	the following inequality holds
	\begin{equation}
	\dot{V}(t)+e^{T}(t)e(t)-\gamma^{2}w^{T}(t)w(t)< 0.
	\end{equation}
	Further, we can derive
	\begin{equation} \label{Hperformance}
	\int_{0}^{L}(\Vert e(t)\Vert^{2}-\gamma^{2}\Vert w(t)\Vert^{2})dt+V(t)|_{t=L}-V(t)|_{t=0}\leq 0,
	\end{equation}
	for $V(t)|_{t=0}=0$, and $V(t)|_{t=L}\geq0$. The inequality (\ref{Hperformance}) can be easily converted to inequality (\ref{Hinf definition}), which implies the $H_{\infty}$ performance requirement is satisfied.
	
	In addition, as $\Theta_1(t)+\Theta_2$ can also be presented as
	\begin{equation}\label{elimination}
	\begin{aligned}
	&\Theta_1(t)+\Theta_2 =hF_{1}O(t)^{-1}F_{1}^{T}+\frac{h}{3}F_{2}O(t)^{-1}F_{2}^{T}\\
	&+\frac{h}{5}F_{3}O(t)^{-1}F_{3}^{T}+Sym\{F_{1}\Pi_{1}+F_{2}\Pi_{2}+F_{3}\Pi_{3}\},
	\end{aligned}
	\end{equation}    
	to reduce computational complexity,  we assume 
	\begin{equation}
	\begin{aligned}
	& F_1=\frac{1}{h}\left[\begin{matrix}
	-O(t) & O(t) & 0 & 0 & 0
	\end{matrix}\right]^{T},\\
	& F_2=\frac{3}{h}\left[
	\begin{matrix}
	-O(t) & -O(t) & 2O(t) & 0 & 0
	\end{matrix}\right]^{T},\\
	& F_3=\frac{5}{h}\left[
	\begin{matrix}
	-O(t) & O(t) & 6O(t) & -60(t) & 0
	\end{matrix}
	\right]^{T}.
	\end{aligned}
	\end{equation}
	Thus, $\Theta_1(t)+\Theta_2$ can be denoted as 
	\begin{equation}
	\Theta_1(t)+\Theta_2=\Xi_1(t)+\Xi_2(t)+\Xi_3(t),
	\end{equation}
	where $\Xi_1(t)$, $\Xi_2(t)$, $\Xi_3(t)$ are defined in (\ref{lemma1}).
	
	According to the Schur Complement criterion, we can easily convert the inequality (\ref{transfer}) as (\ref{lemma1}). Then following similar deduction process, as $w(t)\equiv0$, and in the light of the inequalities (\ref{transfer}), (\ref{deduction}), we can derive $\dot{V}(t)<0$, which means the filtering system (\ref{Closedfiltersystem}) is asymptotically stable. Hence the proof of Lemma 2 is finished.
	\subsubsection*{Remark 1}  From the proof process of Lemma 2, we can see that a new integral inequality (\ref{lemma})  is used to derive the upper bounds of $\dot{V}(t)$, usually the work is finished by the inequalities derived from the Leibniz-Newton formula, compared with these inequalities, the adopted inequality (\ref{lemma}) is tighter, and thus less conservative result can be obtained. Besides, the LMIs (\ref{lemma1}) derived from (\ref{lemma}) is simpler, which means it will be more realizable in practice.
	
	\subsubsection*{Remark 2} From (\ref{LKFfuzzy}), it can be seen that a fuzzy Lyapunov function is adopted, such design will increase the number of the free-weighting matrices, and thus can relax corresponding result. 
	
	\subsubsection*{Lemma 3}
	For system (\ref{Closedfiltersystem}), the constants $h$, $\rho$, $\upsilon$ and $\gamma>0$ are prescribed, it will be asymptotically stable with $w(t)\equiv0$, and satisfy the $H_{\infty}$ performance condition (\ref{Hinf definition}),  if there exist matrices 
	\begin{equation}
	\tilde{M}=\left[\begin{matrix}
	M_{11} & \tilde{M}_{22}\\
	* & \tilde{M}_{22}
	\end{matrix}\right]>0,
	\end{equation} $\tilde{N}(t)\in\mathbb{R}^{2n\times 2n}>0$, $\tilde{O}(t)\in\mathbb{R}^{2n\times 2n}>0$, such that the following LMI is feasible.
	\begin{equation}\label{lemma3}
	\tilde{\Omega}(t)=\left[\begin{matrix}
	\tilde{\Theta}(t) & \sqrt{h}\tilde{\Gamma}^{T}_{1}(t) & \tilde{\Gamma}^{T}_2(t)\\
	* & -2\upsilon \tilde{M}+\upsilon^{2}\tilde{O}(t) & 0\\
	* & * & -1
	\end{matrix}\right]<0,
	\end{equation}
	where
	\begin{flalign*}
	& \tilde{\Theta}(t)=(\tilde{\Xi}_1(t)+\tilde{\Xi}_2(t)+\tilde{\Xi}_3(t))+\tilde{\Theta}_3(t),\\
	&\tilde{\Xi}_1(t)=\frac{1}{h}\left[
	\begin{matrix}
	-\tilde{O}(t) & \tilde{O}(t) & 0 & 0 & 0\\
	* & -\tilde{O}(t) & 0 & 0 & 0\\
	* & * & 0 & 0 & 0\\
	* & * & * & 0 & 0\\
	* & * & * & * & 0
	\end{matrix}\right],&\\
	&\tilde{\Xi}_2(t)=\frac{3}{h}\left[
	\begin{matrix}
	-\tilde{O}(t) & -\tilde{O}(t) & 2\tilde{O}(t) & 0 & 0\\
	* & -\tilde{O}(t) & 2\tilde{O}(t) & 0 & 0\\
	* & * & -4\tilde{O}(t) & 0 & 0\\
	* & * & * & 0 & 0\\
	* & * & * & * & 0
	\end{matrix}\right],&	
	\end{flalign*}
	\begin{flalign*}
	&\tilde{\Xi}_3(t)=\frac{5}{h}\left[
	\begin{matrix}
	-\tilde{O}(t) & \tilde{O}(t) & 6\tilde{O}(t) & -6\tilde{O}(t) & 0\\
	* & -\tilde{O}(t) & -6\tilde{O}(t) & 6\tilde{O}(t) & 0\\
	* & * & -36\tilde{O}(t) & 36\tilde{O}(t) & 0\\
	* & * & * & -36\tilde{O}(t) & 0\\
	* & * & * & * & 0
	\end{matrix}\right],&\\	
	&\tilde{\Theta}_{3}(t)=\left[\begin{matrix}
	Sym\{\lambda_1(t)\}+\tilde{N}(t) & \lambda_2(t) & 0 & 0 & \lambda_3(t)\\
	* & -\tilde{N}(t) & 0 & 0 & 0\\
	* & * & 0 & 0 & 0\\
	* & * & * & 0 & 0\\
	* & * & * & * & -\gamma^{2}
	\end{matrix}\right],	\\	
	&\tilde{\Gamma}_1(t)=\left[\begin{matrix}
	\lambda_1(t) & \lambda_2(t) & 0 & 0 & \lambda_3(t)
	\end{matrix}\right],\\
	&\tilde{\Gamma}_{2}(t)=\left[
	\begin{matrix}
	\left[\begin{matrix}
	E(t) & -\mathscr{C}(t)
	\end{matrix}\right] & \left[\begin{matrix}
	E_{\tau}(t) & 0
	\end{matrix}\right] & 0 & 0 & 0
	\end{matrix}\right],&\\
	&\lambda_1(t)=\left[\begin{matrix}
	M_{11}A(t)+\mathscr{B}(t)C(t) & \mathscr{A}(t)\\
	\tilde{M}_{22}A(t)+\mathscr{B}(t)C(t) & \mathscr{A}(t)
	\end{matrix}\right],&\\
	&\lambda_2(t)=\left[\begin{matrix}
	M_{11}A_\tau(t)+\mathscr{B}(t)C_\tau(t) & 0\\
	\tilde{M}_{22}A_\tau(t)+\mathscr{B}(t)C_\tau(t) & 0
	\end{matrix}\right],&\\
	&\lambda_3(t)=\left[\begin{matrix}
	M_{11}B(t)+\mathscr{B}(t)D(t)\\
	\tilde{M}_{22}B(t)+\mathscr{B}(t)D(t) 
	\end{matrix}\right],&
	\end{flalign*}
	and in this case, a feasible $H_{\infty}$ fileter can be presented as
	\begin{equation}\label{parameter1}
	A'_f(t)=\tilde{M}^{-1}_{22}\mathscr{A}(t),\quad B'_f(t)=\tilde{M}^{-1}_{22}\mathscr{B}(t),\quad C'_f(t)=\mathscr{C}(t).
	\end{equation}	    
	\subsubsection*{proof}		
	For any scalar $\upsilon$, the following inequality is true.
	\begin{equation}\label{omega}
	(\upsilon O(t)-M)O(t)^{-1}(\upsilon O(t)-M)\geq 0,
	\end{equation}	
	and it can also be denoted as
	\begin{equation}
	-MO(t)^{-1}M\leq-2\upsilon M+\upsilon^{2}O(t).
	\end{equation}
	Consequently, if the inequality (\ref{lemma2up}) holds, the inequality (\ref{lemma1}) will be true.
	\begin{equation}\label{lemma2up}
	\left[\begin{matrix}
	\Xi(t) & \sqrt{h}\Gamma^{T}_{1}M & \Gamma^{T}_2\\
	* & -2\upsilon M+\upsilon^{2}O(t) & 0\\
	* & * & -1
	\end{matrix}\right]<0.
	\end{equation}
	Then we'll introduce a partition as
	\begin{equation}
	M=\left[
	\begin{array}{cc}
	M_{11}&M_{12}\\
	*&M_{22}
	\end{array}
	\right],
	\end{equation}
	where  $M_{11}>0$, $M_{22}>0$, and $M_{12}$ is assumed to be invertible via invoking small perturbation if it is necessary.
	
	Assume
	\begin{equation}
	\mathscr{S}=\left[
	\begin{matrix}
	I & 0\\
	* & M^{-T}_{22}M^{T}_{12}
	\end{matrix}\right],
	\end{equation}
	and $\mathscr{R}=diag\{\mathscr{S},\mathscr{S},\mathscr{S},\mathscr{S},1\}$. Pre and post multiplying (\ref{lemma2up}) with $diag\{\mathscr{R},\mathscr{S},1\}$, then the LMI (\ref{lemma3}) can be obtained with the changes of variables as	
	\begin{equation}\label{change}
	\begin{aligned}
	&\tilde{M}_{22}=M_{12}M^{-1}_{22}M^{T}_{12},\quad \tilde{M}=\mathscr{S}^{T}M\mathscr{S}=\left[\begin{matrix}
	M_{11} & \tilde{M}_{22}\\
	* & \tilde{M}_{22}
	\end{matrix}\right]&\\
    & \tilde{Y}=\mathscr{S}^{T}Y\mathscr{S},\quad \tilde{O}(t)=\mathscr{S}^{T}O(t)\mathscr{S}, &\\
	&\mathscr{A}(t)=M_{12}\bar{A}(t)M^{-T}_{22}M^{T}_{12},\quad \mathscr{B}(t)=M_{12}\bar{B}(t),\\
	& \mathscr{C}(t)=\bar{C}(t)M^{-T}_{22}M^{T}_{12}.&
	\end{aligned}
	\end{equation}	
	From (\ref{change}), we can obtain:
	\begin{equation}
	\begin{aligned}
	&A_f(t)=M_{12}^{-1}\mathscr{A}(t)M^{-T}_{12}M^{T}_{22},\quad B_f(t)=M_{12}^{-1}\mathscr{B}(t),\\
	&C_f(t)=\mathscr{C}(t)M_{12}^{-T}M_{22}^{T}.
	\end{aligned}
	\end{equation}
	
	Besides, as $\tilde{M}_{22}=M_{12}M^{-1}_{22}M^{T}_{12}$, through an equivalent transformation $M_{12}^{-T}M_{22}x_f(t)$, we can obtain an admissible fuzzy $H_{\infty}$ realization as:
	\begin{equation}
	\begin{aligned}
	&A'_f(t)=M_{12}^{-T}M_{22}(M_{12}^{-1}\mathscr{A}(t)M^{-T}_{12}M^{T}_{22})M_{22}^{-T}M^{T}_{12}=\tilde{M}^{-1}_{22}\mathscr{A}(t),\\ &B'_f(t)=M_{12}^{-T}M_{22}(M_{12}^{-1}\mathscr{B}(t))=\tilde{M}^{-1}_{22}\mathscr{B}(t),\\
	& C'_f(t)=(\mathscr{C}(t)M_{12}^{-T}M_{22}^{T})M_{22}^{-T}M^{T}_{12}=\mathscr{C}(t).
	\end{aligned}
	\end{equation}
Thus, we complete the proof of Lemma 3.

Lemma 3 provides a feasible $H_{\infty}$ filter for system (\ref{fuzzy model}), however, it is necessary to point out that the equation (\ref{parameter1}) cannot be directly applied to fuzzy filter design. To address this problem, we will transfer the conditions in Lemma 3 into a finite set of LMIs.

	\subsubsection*{Theorem 1}
	For system (\ref{Closedfiltersystem}), the constants $h$, $\rho$, $\upsilon$ and $\gamma>0$ are prescribed, it will be asymptotically stable with $w(t)\equiv0$, and satisfy the $H_{\infty}$ performance condition (\ref{Hinf definition}),  if there exist matrices 
	\begin{equation}
	\tilde{M}=\left[\begin{matrix}
	M_{11} & \tilde{M}_{22}\\
	* & \tilde{M}_{22}
	\end{matrix}\right]>0,
	\end{equation} $\tilde{N}_i\in\mathbb{R}^{2n\times 2n}>0$, $\tilde{O}_i\in\mathbb{R}^{2n\times 2n}>0$,  such that the following LMIs are feasible.
	\begin{equation}\label{Theorem 1}
	\tilde{\Omega}_{ij}=\left[\begin{matrix}
	\tilde{\Theta}_{ij} & \sqrt{h}\tilde{\Gamma}^{T}_{1ij} & \tilde{\Gamma}^{T}_{2ij}\\
	* & -2\upsilon \tilde{M}+\upsilon^{2}\tilde{O}_i & 0\\
	* & * & -1
	\end{matrix}\right]<0
	\end{equation}
	where
	\begin{flalign*}
	& \tilde{\Theta}_{ij}=(\tilde{\Xi}_{1i}+\tilde{\Xi}_{2i}+\tilde{\Xi}_{3i})+\tilde{\Theta}_{3ij},\\
	&\tilde{\Xi}_{1i}=\frac{1}{h}\left[
	\begin{matrix}
	-\tilde{O}_i & \tilde{O}_i & 0 & 0 & 0\\
	* & -\tilde{O}_i & 0 & 0 & 0\\
	* & * & 0 & 0 & 0\\
	* & * & * & 0 & 0\\
	* & * & * & * & 0
	\end{matrix}\right],&\\
	&\tilde{\Xi}_{2i}=\frac{3}{h}\left[
	\begin{matrix}
	-\tilde{O}_i & -\tilde{O}_i & 2\tilde{O}_i & 0 & 0\\
	* & -\tilde{O}_i & 2\tilde{O}_i & 0 & 0\\
	* & * & -4\tilde{O}_i & 0 & 0\\
	* & * & * & 0 & 0\\
	* & * & * & * & 0
	\end{matrix}\right],&\\
	&\tilde{\Xi}_{3i}=\frac{5}{h}\left[
	\begin{matrix}
	-\tilde{O}_i & \tilde{O}_i & 6\tilde{O}_i & -6\tilde{O}_i & 0\\
	* & -\tilde{O}_i & -6\tilde{O}_i & 6\tilde{O}_i & 0\\
	* & * & -36\tilde{O}_i & 36\tilde{O}_i & 0\\
	* & * & * & -36\tilde{O}_i & 0\\
	* & * & * & * & 0
	\end{matrix}\right],&\\
	&\tilde{\Theta}_{3ij}=\left[\begin{matrix}
	Sym\{\lambda_{1ij}\}+\tilde{N}_i & \lambda_{2ij} & 0 & 0 & \lambda_{3ij}\\
	* & -\tilde{N}_i & 0 & 0 & 0\\
	* & * & 0 & 0 & 0\\
	* & * & * & 0 & 0\\
	* & * & * & * & -\gamma^{2}
	\end{matrix}\right]\\
	&\tilde{\Gamma}_{1ij}=\left[\begin{matrix}
	\lambda_{1ij} & \lambda_{2ij} & 0 & 0 & \lambda_{3ij}
	\end{matrix}\right]\\
	&\tilde{\Gamma}_{2ij}=\left[
	\begin{matrix}
	\left[\begin{matrix}
	E_i & -\mathscr{C}_j
	\end{matrix}\right] & \left[\begin{matrix}
	E_{\tau i} & 0
	\end{matrix}\right] & 0 & 0 & 0
	\end{matrix}\right],&\\
	&\lambda_{1ij}=\left[\begin{matrix}
	M_{11}A_i+\mathscr{B}_jC_i & \mathscr{A}_j\\
	\tilde{M}_{22}A_i+\mathscr{B}_jC_i & \mathscr{A}_j
	\end{matrix}\right]&	
	\end{flalign*}
	\begin{flalign*}
	&\lambda_{2ij}=\left[\begin{matrix}
	M_{11}A_{\tau i}+\mathscr{B}_jC_{\tau i} & 0\\
	\tilde{M}_{22}A_{\tau i}+\mathscr{B}_jC_{\tau i} & 0
	\end{matrix}\right],&\\
	&\lambda_{3ij}=\left[\begin{matrix}
	M_{11}B_i+\mathscr{B}_jD_i\\
	\tilde{M}_{22}B_i+\mathscr{B}_iD_i
	\end{matrix}\right],&\\
	& i=1,2,...,p, j=1,2,...,c.    
	\end{flalign*}
	and in this case, the parameters of the fuzzy filter can be presented as
	\begin{equation}\label{parameter}
	A'_{fj}=\tilde{M}^{-1}_{22}\mathscr{A}_j,\quad B'_{fj}=\tilde{M}^{-1}_{22}\mathscr{B}_j,\quad C'_{fj}=\mathscr{C}_j.
	\end{equation}	        
	\subsubsection*{proof} 
	Since
	\begin{equation}
	\sum_{i=1}^{p}\phi_i(\psi(t))=\sum_{j=1}^{c}n_j(g(t))=\sum_{i=1}^{p}\sum_{j=1}^{c}\phi_i(\psi(t))n_j(g(t))=1,
	\end{equation}
	and in terms of (\ref{simplication}) and (\ref{fuzzy}), we can derive
	\begin{equation}
	\tilde{\Omega}(t)=\sum_{i=1}^{p}\sum_{j=1}^{c}\phi_i(\psi(t))n_j(g(t))\tilde{\Omega}_{ij},
	\end{equation}
	where $\tilde{\Omega}_{ij}$ is defined in (\ref{Theorem 1}).
	
	As a result, if $\tilde{\Omega}_{ij}<0$ holds, $\tilde{\Omega}(t)<0$ can be derived. And according to Lemma 3, we know that $\tilde{\Omega}(t)<0$ means the filtering system (\ref{Closedfiltersystem}) is asymptotically stable and satisfy the $H_{\infty}$ performance condition. Hence, the proof of Theorem 1 is accomplished.		
	\subsubsection*{Theorem 2}
	For system (\ref{Closedfiltersystem}), the constants $h$, $\rho$, $\upsilon$ and $\gamma>0$ are prescribed, it will be asymptotically stable with $w(t)\equiv0$, and satisfy the $H_{\infty}$ performance condition (\ref{Hinf definition}),  if there exist matrices 
	\begin{equation}
	\tilde{M}=\left[\begin{matrix}
	M_{11} & \tilde{M}_{22}\\
	* & \tilde{M}_{22}
	\end{matrix}\right]>0,
	\end{equation} $\tilde{N}_i\in\mathbb{R}^{2n\times 2n}>0$, $\tilde{O}_i\in\mathbb{R}^{2n\times 2n}>0$, $M_{ij}\in\mathbb{R}^{(10n+2)\times(10n+2)}$, $M_{ij}\in\mathbb{R}^{(10n+2)\times(10n+2)}$, such that the following LMIs are feasible.
	\begin{equation}\label{Theorem 2}
	\begin{aligned}
	&\tilde{\Omega}_{ij}-M_{ij}+Q_{ij}+\sum_{r=1}^{p}\sum_{s=1}^{c}\bar{d}_{ij}M_{rs}-\sum_{a=1}^{p}\sum_{b=1}^{c}\underline{d}_{ij}M_{ab}<0,\\
	& i=1,2,...,p, j=1,2,...,c,
	\end{aligned}
	\end{equation}
	where $\tilde{\Omega}_{ij}$ is defined in (\ref{Theorem 1}). And in this case, the parameters of the fuzzy filter can also be presented as (\ref{parameter}).            
	\subsubsection*{proof}    
	In this part, for the convenience of notations, we denote
	\begin{equation}
	\phi_i(\psi(t))=\phi_i,\quad n_j(g(t))=n_j,\quad\phi_i(\psi(t))n_j(g(t))=d_{ij},
	\end{equation}       
	and assume $\underline{d}_{ij}$ and $\bar{d}_{ij}$ are the lower bound and upper bound of $d_{ij}$, respectively.
	
	From Lemma 2 and Lemma 3, we have
	\begin{equation}
	\begin{aligned}
	&\dot{V}(t)+e^{T}(t)e(t)-\gamma^{2}w^{T}(t)w(t)<\xi^{T}(t)\tilde{\Omega}(t)\xi(t),
	\end{aligned}
	\end{equation}
	so we can derive
	\begin{equation}
	\begin{aligned}
	&\xi^{T}(t)\tilde{\Omega}(t)\xi(t)\\
	&=\sum_{i=1}^{p}\sum_{j=1}^{c}d_{ij}\xi^{T}(t)\Omega_{ij}\xi(t)\\
	&\leq \sum_{i=1}^{p}\sum_{j=1}^{c}d_{ij}\xi^T(t)\Omega_{ij}\xi(t) + \sum_{i=1}^{p}\sum_{j=1}^{c}(\bar{d}_{ij}-d_{ij})\xi^T(t)M_{ij}\xi(t)\\
	&\quad+\sum_{i=1}^{p}\sum_{j=1}^{c}(d_{ij}-\underline{d}_{ij})\xi^T(t)Q_{ij}\xi(t)\\
	&=\sum_{i=1}^{p}\sum_{j=1}^{c}d_{ij}\xi^{T}(t)(\Omega_{ij}-M_{ij}+Q_{ij})\xi(t)\notag	
	\end{aligned}
	\end{equation}
	\begin{equation}
	\begin{aligned}
	&\quad+\sum_{i=1}^{p}\sum_{j=1}^{c}\bar{d}_{ij}\xi^{T}(t)M_{ij}\xi(t)-\sum_{i=1}^{p}\sum_{j=1}^{c}\underline{d}_{ij}\xi^{T}(t)Q_{ij}\xi(t)\\
	&=\sum_{i=1}^{p}\sum_{j=1}^{c}d_{ij}\xi^{T}(t)(\Omega_{ij}-M_{ij}+Q_{ij}+\sum_{r=1}^{p}\sum_{s=1}^{c}\bar{d}_{rs}M_{rs}\\
	&\quad-\sum_{a=1}^{p}\sum_{b=1}^{c}\underline{d}_{ab}Q_{ab})\xi(t).
	\end{aligned}
	\end{equation}
	Consequently, if LMIs (\ref{Theorem 2}) holds, we can get $\tilde{\Omega}(t)<0$, which means both the $H_{\infty}$ performance condition (\ref{Hinf definition}) and the asymptotically stable requirement can be satisfied. Thus, the proof of Theorem 2 is finished.
	
	\subsection*{Remark 3} It can be seen that Theorem 2 includes the information of the membership functions while Theorem 1 does not. As a result, Theorem 2 is less conservative than Theorem 2. Whereas, on the other hand, Theorem 2 also includes complex matrices $M_{ij}$, $Q_{ij}$, $(i=1,...p, j=1,...,c)$, which means it will be more difficult to implement in engineering applications. So both Theorem 1 and Theorem 2 have their own significance in practice.

	\section{Simulation}	
	In this section, two simulation examples will be provided to demonstrate the effectiveness and superiority of the designed criteria.
	\subsection{Example 1}
	Consider a time-delay system in (\ref{fuzzy model}) with
	\begin{flalign*}
	&A_{1}=\left[\begin{matrix}
	-2.1 & 0.1\\ 1 & -2 
	\end{matrix}\right],
	A_{2}=\left[\
	\begin{matrix}
	-1.1 & 0\\ -0.2 & -1.1
	\end{matrix}\right],&\\
	&A_{\tau 1}=\left[
	\begin{matrix}
	-1.1 & 0.1 \\ -0.8 & -0.9
	\end{matrix}\right],
	A_{\tau 2}=\left[
	\begin{matrix}
	-0.9 & 0 \\ -1.1 & -1.2
	\end{matrix}\right],&\\
	&B_{1}=\left[\
	\begin{matrix}
	1\\-0.2
	\end{matrix}\right],
	B_{2}=\left[\
	\begin{matrix}
	0.3\\0.1
	\end{matrix}\right],&\\
	&C_{1}=\left[\
	\begin{matrix}
	1 & 0
	\end{matrix}\right],
	C_{2}=\left[\
	\begin{matrix}
	0.5 & -0.6
	\end{matrix}\right],&\\
	&C_{\tau 1}=\left[\
	\begin{matrix}
	-0.8 & 0.6
	\end{matrix}\right],
	C_{\tau 2}=\left[\
	\begin{matrix}
	-0.2 & 1
	\end{matrix}\right],&\\
	& D_1=0.3, D_2=-0.6, &\\
	&E_{1}=\left[\
	\begin{matrix}
	1 & -0.5
	\end{matrix}\right],
	E_{2}=\left[\
	\begin{matrix}
	-0.2 & 0.3
	\end{matrix}\right],&\\
	&E_{\tau 1}=\left[\
	\begin{matrix}
	0.1 & 0
	\end{matrix}\right],
	E_{\tau 2}=\left[\
	\begin{matrix}
	0 & 0.2
	\end{matrix}\right].&\\	
	\end{flalign*}
	and the membership functions of the fuzzy model and the fuzzy filter are chosen as
	\begin{flalign*}
	&\phi_{1}(\psi(t))=1-\frac{0.5}{1+e^{-3-t}},&\\
	&\phi_{2}(\psi(t))=1-\phi_{1}(\psi(t)),&\\
	& n_{1}(g(t))= 0.7-\frac{0.5}{1+e^{4-t}},&\\
	& n_{2}(g(t))=1-n_{1}(g(t)).&	
	\end{flalign*}	
	Assume $(\rho,\upsilon,h)=(0.2,1,0.5)$, using the LMIs (\ref{Theorem 2}) presented in Theorem 2, the minimum attenuation level $\gamma=0.18$ can be acquired, and corresponding feasible solutions are as follows.
	\begin{equation*}
	\begin{aligned}
	&\tilde{M}_{22}=\left[\begin{matrix}
	0.1000 &  -0.0039\\
	-0.0039 &   0.1696
	\end{matrix}\right],\\
	&\mathscr{A}_{1}=\left[\begin{matrix}
	-0.3809 &  -0.0041\\
	0.0639  & -0.3911 
	\end{matrix}\right],\\
	&\mathscr{A}_{2}=\left[\begin{matrix}
	-0.2370  &   0.0922\\
	0.1954  & -0.0251
	\end{matrix}\right],  \\
	&\mathscr{B}_{1}=\left[\begin{matrix}
	-0.2688\\
	-0.0406
	\end{matrix}\right],  
	\mathscr{B}_{2}=\left[\begin{matrix}
	-0.1122\\
	0.0642
	\end{matrix}\right],  \\  
	&\mathscr{C}_{1}=\left[\begin{matrix}
	-0.5095  &  0.1230  	
	\end{matrix}\right],\\
	&\mathscr{C}_{2}=\left[\begin{matrix}
	-0.2762  &  0.0016
	\end{matrix}\right].
	\end{aligned}
	\end{equation*}	  
	Further, applying (\ref{parameter}), we can get the parameters of the $H_{\infty}$ filter as:
	\begin{equation*}
	\begin{aligned}
	& A_{f1}=\left[\begin{matrix}
	-3.7967 &  -0.1318\\
	0.2891  & -2.3097
	\end{matrix}\right],\\
	& A_{f2}=\left[\begin{matrix}
	-2.3263 &   0.9168\\
	1.0983 &  -0.1267
	\end{matrix}\right],\\
	& B_{f1}=\left[\begin{matrix}
	-2.6985\\
	-0.3019
	\end{matrix}\right],
	B_{f2}=\left[\begin{matrix}
	-1.1075\\
	0.3529
	\end{matrix}\right],\\
	& C_{f1}=\left[\begin{matrix}
	-0.5095  &  0.1230  
	\end{matrix}\right],\\
	& C_{f 2}=\left[\begin{matrix}
	-0.2762  &  0.0016
	\end{matrix}\right].
	\end{aligned}
	\end{equation*}	
	As we have noted the fact that different $(h,\upsilon)$ will produce different minimum attenuation level $\gamma$, and to fully demonstrate the validity of the proposed method, we will respectively apply Theorem 2 in this paper and the methods presented in \cite{star,Y,JH,short,An2015} to compute the minimum attenuation level $\gamma$ with different $(h,\upsilon)$. And the corresponding results are listed in Table \ref{tb1}-\ref{tb3}. 	
	\begin{table}[!h]
		\caption{ The minimum attenuation level $\gamma$ for $\upsilon=2$}\label{tb1}
		\centering
		\begin{tabular}{c c c c c c c }
			\hlinew{0.75pt}
			method & $h=0.5$ & $h=0.6$ & $h=0.8$ & $h=1$ \\
			\hline  			
			\cite{star} & 0.35 & 0.36 & 0.38 & 0.41\\
			\cite{Y} & 0.25 & 0.25 & 0.27 & 0.29\\
			\cite{JH} & 0.25 & 0.25 & 0.27 & 0.29\\
			\cite{short} & 0.24 & 0.25 & 0.25 & 0.26\\
			\cite{An2015} & 0.23 & 0.24 & 0.25 & 0.25\\
			Th. 2 & 0.17 & 0.17 & 0.17 & 0.18\\
			\hlinew{0.75pt}
		\end{tabular}
	\end{table}  	
	
	\begin{table}[!h]
		\caption{ The minimum attenuation level $\gamma$ for $\upsilon=5$}\label{tb2}
		\centering
		\begin{tabular}{c c c c c c c }
			\hlinew{0.75pt}
			method & $h=0.5$ & $h=0.6$ & $h=0.8$ & $h=1$ \\
			\hline  			
			\cite{star} & 0.34 & 0.34 & 0.35 & 0.37\\
			\cite{Y} & 0.24 & 0.24 & 0.25 & 0.26\\
			\cite{JH} & 0.24 & 0.24 & 0.25 & 0.26\\
			\cite{short} & 0.24 & 0.24 & 0.25 & 0.26\\
			\cite{An2015} & 0.23 & 0.24 & 0.24 & 0.25\\
			Th. 2 & 0.16 & 0.17 & 0.17 & 0.17\\
			\hlinew{0.75pt}
		\end{tabular}
	\end{table}  
	
	\begin{table}[!h]
		\caption{ The minimum attenuation level $\gamma$ for $\upsilon=20$}\label{tb3}
		\centering
		\begin{tabular}{c c c c c c c }
			\hlinew{0.75pt}
			method & $h=0.5$ & $h=0.6$ & $h=0.8$ & $h=1$ \\
			\hline  			
			\cite{star} & 0.37 & 0.45 & 1.01 & $--$\\
			\cite{Y} & 0.26 & 0.32 & 0.70 & $--$\\
			\cite{JH} & 0.26 & 0.28 & 0.44 & $--$\\
			\cite{short} & 0.25 & 0.26 & 0.35 & 0.45\\
			\cite{An2015} & 0.23 & 0.24 & 0.25 & 0.25\\
			Th. 2 & 0.17 & 0.17 & 0.17 & 0.17\\
			\hlinew{0.75pt}
		\end{tabular}
	\end{table}   		  			
	
	where $--$ denotes that the minimum attenuation level $\gamma$ does not exist. 	
	
	From Table \ref{tb1}-\ref{tb3}, we can clearly see that smaller minimum attenuation level $\gamma$ can be obtained with Theorem 3 than the ones obtained with other methods, which implies that the method proposed in this paper is less conservative than those in (\cite{Y,short,JH,star,An2015}).
	
	\subsubsection*{Remark 4} The less conservative results can be obtained with the approach proposed in this paper mainly because of three reasons. First, the new integral inequality (\ref{lemma}) is employed to derive stability condition, which is tighter than those derived from the Leibniz-Newton formula. Besides, our criterion is membership functions dependent while the others are membership functions independent. Also, we adopt fuzzy Lyapunov function, we can increase the number of free-weighting matrices, and further relax the result.

	\subsection{Example 2}	
	Consider a system in the form of (\ref{fuzzy model}) with
	\begin{flalign*}
	& A_{1}=\left[\begin{matrix}
	-2.1 & 0.1\\ 1 & -2 
	\end{matrix}\right],
	A_{2}=\left[\
	\begin{matrix}
	-1.9 & 0\\ -0.2 & -1.1
	\end{matrix}\right],
	A_{3}=\left[\
	\begin{matrix}
	-1 & 0 \\ 1 & 0
	\end{matrix}\right],&\\
	&A_{\tau 1}=\left[
	\begin{matrix}
	-1.1 & 0.1 \\ -0.8 & -0.9
	\end{matrix}\right],
	A_{\tau 2}=\left[
	\begin{matrix}
	-0.9 & 0 \\ -1.1 & -1.2
	\end{matrix}\right],&\\  
	& A_{\tau 3}=\left[
	\begin{matrix}
	-1 & 0 \\ 0.5 & 0
	\end{matrix}\right],
	B_{1}=\left[\
	\begin{matrix}
	1\\-0.2
	\end{matrix}\right],
	B_{2}=\left[\
	\begin{matrix}
	0.3\\0.1
	\end{matrix}\right],
	B_{3}=\left[\
	\begin{matrix}
	0.5\\-1
	\end{matrix}\right],&\\
	&C_{1}=\left[\
	\begin{matrix}
	1 & 0
	\end{matrix}\right],
	C_{2}=\left[\
	\begin{matrix}
	0.5 & -0.6
	\end{matrix}\right],
	C_{3}=\left[\
	\begin{matrix}
	0.5 & -0.6
	\end{matrix}\right],&\\
	&C_{\tau 1}=\left[\
	\begin{matrix}
	-0.8 & 0.6
	\end{matrix}\right]\quad
	C_{\tau 2}=\left[\
	\begin{matrix}
	-0.2 & 1
	\end{matrix}\right],
	C_{\tau 3}=\left[\
	\begin{matrix}
	0 & 0.5
	\end{matrix}\right],&\\
	& D_{1}=0.3, D_{2}=-0.6, D_{3}=0.5&\\
	&E_{1}=\left[\
	\begin{matrix}
	1 & -0.5
	\end{matrix}\right],
	E_{2}=\left[\
	\begin{matrix}
	-0.2 & 0.3
	\end{matrix}\right],
	E_{3}=\left[\
	\begin{matrix}
	-0.6 & 1
	\end{matrix}\right],&\\
	&E_{\tau 1}=\left[\
	\begin{matrix}
	0.1 & 0
	\end{matrix}\right],
	E_{\tau 2}=\left[\
	\begin{matrix}
	0 & 0.2
	\end{matrix}\right],
	E_{\tau 3}=\left[\
	\begin{matrix}
	0.2 & 0.1
	\end{matrix}\right].&
	\end{flalign*}
	and the membership functions are defined as
	\begin{flalign*}
	&\phi_{1}(\psi(t))=1-\frac{0.6}{1+e^{-3-t}},&\\
	&\phi_{2}(\psi(t))=\frac{0.4}{1+e^{-3-t}},&\\
	&\phi_{3}(\psi(t))=1-\phi_{1}(\psi(t))-\phi_{2}(\psi(t)),&\\
	& n_{1}(g(t))= 0.7-\frac{0.5}{1+e^{4-t}},\\ & n_{2}(g(t))=1-n_{1}(g(t)).&	
	\end{flalign*}	
	Then Theorem 1 and Theorem 2 will be adopted to compute the minimum attenuation level $\gamma$, the corresponding results are listed in Table \ref{tb4}-\ref{tb5}	
	\begin{table}[!h]
		\renewcommand\arraystretch{1.2}
		\caption{ The minimum attenuation level $\gamma$ for $h=0.5$}\label{tb4}
		\begin{tabular}{c c c c c c c }
			\hlinew{0.75pt}
			method & $\upsilon=0.7$ & $\upsilon=1$ & $\upsilon=2$ & $\upsilon=5$ & $\upsilon=10$ & $\upsilon=20$\\
			\hline  			
			Th. 1 & 0.40 & 0.27 & 0.24 & 0.23 & 0.22 & 0.24\\
			Th. 2 & 0.20 & 0.16 & 0.13 & 0.11 & 0.09 & 0.10\\
			\hlinew{0.75pt}
		\end{tabular}
	\end{table}
	\begin{table}[!h]
		\renewcommand\arraystretch{1.2}
		\caption{ The minimum attenuation level $\gamma$ for $h=0.8$}\label{tb5}
		\begin{tabular}{c c c c c c c }
			\hlinew{0.75pt}
			method & $\upsilon=0.7$ & $\upsilon=1$ & $\upsilon=2$ & $\upsilon=5$ & $\upsilon=10$ & $\upsilon=20$\\
			\hline  			
			Th. 1 & 4.53 & 0.55 & 0.26 & 0.24 & 0.24 & 0.25\\
			Th. 2 & 0.64 & 0.42 & 0.15 & 0.12 & 0.11 & 0.11\\
			\hlinew{0.75pt}
		\end{tabular}
	\end{table}	
	
	From Table \ref{tb4}-\ref{tb5}, we can clearly see that Theorem 2 can produce much smaller minimum attenuation level $\gamma$ than Theorem 1, which means Theorem 2 is more relaxed than Theorem 1. And this is because Theorem 2 takes the information of the membership functions into consideration while Theorem 1 does not. Besides, in the simulation process, Theorem 2 spent much longer time to compute the minimum attenuation level $\gamma$ than Theorem 1, the reason for this phenomenon is that LMIs (\ref{Theorem 1}) in Theorem 1 is simpler than the LMIs (\ref{Theorem 2}) in Theorem 2, which also implies Theorem 1 will be more realizable than Theorem 2 in practice. 

	\subsubsection*{Remark 5}
	It is essential to point out that in Example 2, the fuzzy model has 3 fuzzy rules while the fuzzy filter have 2 fuzzy rules, and the membership functions of the fuzzy model and the fuzzy filter are totally different. Such design means the number of fuzzy rules and the membership functions of the fuzzy filter can be freely chosen, thus, we can lower the implementation cost via choosing simpler membership functions, and successfully deal with the situation with uncertainty through avoiding unknown membership functions. 
	

	\section{Conclusions}
	In this paper, the fuzzy $H_{\infty}$ filter design problem has been investigated for nonlinear time-delay systems. The T-S fuzzy model has been used to describe the dynamics of the system, and two LMI-based criteria have been derived. Unlike conventional fuzzy $H_{\infty}$ filter, The designed fuzzy filter has been allowed to freely choose the premise membership functions and the number of rules, ergo, robustness to uncertainty and lower implementation cost can be realized. Besides, to reduce the conservatism of the derived results, a novel integral inequality which is tighter than other existing ones has been introduced, and a fuzzy Lyapunov function and the information of the membership functions have been taken into account, too.
	Finally, three examples have demonstrated the validity of the designed fuzzy $H_{\infty}$ filter.


\begin{thebibliography}{}  
		\bibitem[Cao and Frank(2000)]{analysis2000}
		Y. Y. Cao, and P. M. Frank.
		\newblock Analysis and synthesis of nonlinear time delay
		systems via fuzzy control approach.
		\newblock \emph{IEEE Transactions on Fuzzy System}, 8(4), 200-211, 2000.
		
		\bibitem[Jeung et al.(1998)Jeung, Kim, and Park]{H1998}
		E. T. Jeung, J.H. Kim and H. B. Park.\newblock $H_{\infty}$-output feedback controller design for linear systems with time-varying delayed state.\newblock \emph{IEEE
			Transactions on Automatic Control}, 43, 971-974, 1998.	
		\bibitem[Wu et~al.(2004)Wu, He ,She, and Liu]{delayt2004}
		M. Wu, Y. He, J. H. She and G. P. Liu.\newblock Delay-dependent criteria for robust stability of time-varying delay systems.\newblock \emph{Automatica}, 40(8), 1435-1439, 2004.
		
        \bibitem[Kong and Dai(2016)]{kong2016new}
        L. Kong, R. Dai and Y. Zhang.\newblock A new quality model for object detection using compressed videos.\newblock \emph{2016 IEEE International Conference on Image Processing (ICIP)}, 3797-3801, 2016.
		
		\bibitem[Bernstein and Haddad(1989)]{steady1989}
		D. S. Bernstein and W. M. Haddad.\newblock Steady-state Kalman filtering with
		an $H_{\infty}$ error bound.\newblock \emph{Systems $\&$ Cotrol Letters}, 12, 9-16, 1989.
		
		
		\bibitem[Gao and Wang(2003)]{Delay2003}
		H. Gao and C. Wang.\newblock Delay-dependent robust $H_{\infty}$ and
		$L_{2}$$-$$L_{\infty}$ filtering for a class of uncertain nonlinear time-delay systems.\newblock \emph{IEEE Transactions on Automatic Control}, 48(9), 1661-1666, 2003.
		
        \bibitem[Kong and Dai(2017)]{kong2017object}
        L. Kong and R. Dai.\newblock Object-detection-based video compression for wireless surveillance systems.\newblock \emph{IEEE MultiMedia}, 24(2), 76-85, 2017.

        \bibitem[Kong and Dai(2016)]{kong2016temporal}
        L. Kong and R. Dai.\newblock Temporal-fluctuation-reduced video encoding for object detection in wireless surveillance systems.\newblock \emph{2016 IEEE International Symposium on Multimedia (ISM)}, 126--132, 2016.
		
		
		\bibitem[Fridman and Shaked(2001)]{A2001}
		E. Fridman and U. Shaked.\newblock A new $H_{\infty}$ filter design for linear time delay systems.\newblock \emph{IEEE Transactions on Signal Processing}, 49(11), 2839-2843, 2001.
		
		
		\bibitem[Tseng and Chen(2001)]{fuzzy2001}
		C. S. Tseng and B. S. Chen.\newblock Fuzzy estimation for a class of nonlinear
		discrete-time dynamic systems.\newblock \emph{IEEE Transactions on Signal Processing},
		49(22), 2605-2619, 2001.
		
        \bibitem[Kong and Dai(2016)]{kong2017impact}
        L. Kong.\newblock Impact of distributed caching on video streaming quality in information centric networks.\newblock \emph{2017 IEEE International Symposium on Multimedia (ISM)}, 399-402, 2017.
		
		\bibitem[Huang et~al.(2011)Huang, He and Zhang]{short}
		S. J. Huang, X. Q. He and N. N. Zhang.\newblock New results on $H_{\infty}$ filter design for nonlinear systems with time delay via T–S fuzzy models.\newblock \emph{IEEE Transactions on Fuzzy Systems}, 19(1), 193-199, 2011.
		
		\bibitem[Lin et~al.(2008)]{star}
		C. Lin, Q. G. Wang, T. H. Lee and B. Chen.\newblock $H_{\infty}$ filter design for nonlinear systems with time-delay through T–S fuzzy model Approach.\newblock \emph{IEEE Transactions on Fuzzy Systems}, 16(13), 739-746, 2008.		
		
		\bibitem[Lam and Narimani(2009)]{imperfect2009}
		H. K. Lam and Mohammad Narimani.\newblock Stability analysis and performance design for fuzzy-model-based control system under imperfect premise matching.\newblock \emph{IEEE Transactions on Fuzzy Systems}, 17(4), 949-961, 2009.
		
		
		\bibitem[Hong et~al.(2015)Hong, Yong, Min and Hua]{New2015}
		B. Z. Hong, H. Yong, W. Min and S. J. Hua.\newblock New results on stability analysis for systems with discrete distributed delay.\newblock \emph{Automatica}, 60, 189-192, 2015.  	
		
		\bibitem[He et~al.(2007)He, Wang, Xie and Lin]{further2007}
		Y. He, Q. G. Wang, L. Xie and C. Lin.\newblock Further improvement of free-weighting matrices technique for systems with time-varying delay. \emph{IEEE Transactions on Automatic Control}, 52(2), 293-299, 2007.	 		  		 
		
		\bibitem[Qiu et~al.(2009)Qiu, Gang and Yang]{new2009}
		J. Qiu, F. Gang and J. Yang.\newblock A new design of delay-dependent robust $ H_{\infty}$
		filtering for discrete-time T–S fuzzy systems
		with time-varying delay.\newblock \emph{International Journal on Fuzzy Systems}, 17(5), 1044-1058, 2009.
		
		
		
		\bibitem[Ma(2022)]{ma2022}	
		Q. Ma.\newblock First-order optimization methods for networked high dimensional systems.\newblock  \emph{Ph.D Thesis, Department of ECE, Boston University}, 2022.
		
		\bibitem[Lin et al.(2007)Lin, Wang, Lee and He]{Fuzzy2007}
				C. Lin, Q. G. Wang, T. H. Lee and Y. He. \newblock Fuzzy weighting-dependent approach to $H_{\infty}$ filter
				design for time-delay fuzzy systems. \newblock \emph{IEEE Transactions on Signal Processing}, 55(6), 2007.
		
		\bibitem[Su et~al.(2009)Su, Chen and Zhang]{Y}
		Y. K. Su, B. Chen, C. Lin, and H. G. Zhang.\newblock A new fuzzy $H_{\infty}$filter
		design for nonlinear continuous-time dynamic systems with time-varying
		delays \emph{Fuzzy Sets and Systems}, 160, 3539-3549, 2009.
		
		
		      		\bibitem[Ma et~al.(2017)]{Ma2017b}	
	Q. Ma, H. Xia, G. Ma, Y. Xia, and C. Wang.\newblock Improved stability and stabilization criteria for TS fuzzy systems with distributed time-delay.\newblock  \emph{International Conference on Data Mining and Big Data}, 517-526, 2017.
		
		\bibitem[Zhang et~al.(2009)Zhang, Xiao and Tao]{JH}
		J. H. Zhang, Y. Q. Xiao, and R. Tao.\newblock New results on $H_{\infty}$ filtering for fuzzy time-delay systems. \emph{IEEE Transactions on Fuzzy System}, 17(1), 128-137, 2009.		
		
		\bibitem[Zhou and He(2015)]{An2015}		
		T. Zhou and X. Q. He.\newblock An improved $H_{\infty}$ filter design for nonlinear systems
		described by T-S fuzzy models with
		time-varying delay.\newblock  \emph{International Journal of Automation and Computing}, 12(6), 671-678, 2015.	
		
		

		\bibitem[Kong and Dai(2018)]{kong2018efficient}	
		L. Kong and R. Dai.\newblock Efficient video encoding for automatic video analysis in distributed wireless surveillance systems.\newblock  \emph{ACM Transactions on Multimedia Computing, Communications, and Applications (TOMM)}, 14(3), 1-24, 2018.
		

		
   
  		\bibitem[Wang et~al.(2020)]{Wang2020}	
		L. Wang, B. Zong, Q. Ma, W. Cheng, J. Ni, W. Yu, Y. Liu, D. Song, H. Chen and Y. Fu.\newblock Inductive and unsupervised representation learning on graph structured object.\newblock  \emph{International conference on learning representations}, 2020.
		
		
  		\bibitem[Qin et~al.(2021)]{Qin2021}	
	C. Qin, L. Wang, Q. Ma, Y. Yin, H. Wang, and Y. Fu.\newblock Contradictory structure learning for semi-supervised domain adaptation.\newblock  \emph{Proceedings of the 2021 SIAM International Conference on Data Mining (SDM)}, 576-584, 2021.	
   
   
   		
  		\bibitem[Ma et~al.(2017)]{Ma2017}	
	Q. Ma, L. Li, J. Shen, H. Guan, G. Ma, and H. Xia.\newblock Improved fuzzy H$_\infty$ filter design method for nonlinear systems with time-varing delay.\newblock  \emph{2017 IEEE International Conference on Systems, Man, and Cybernetics (SMC)}, 722-727, 2017.
	
   
   

		
		
		
		
	\end{thebibliography}
	

\end{document}